# Bed-Load Transport Rate Based on the Entrainment Probabilities of Sediment Grains by Rolling and Lifting

**Jun-De Li***

College of Engineering and Science, Victoria University, Australia, PO Box 14428, MC 8001

Email: Jun-De.Li@vu.edu.au

State Key Laboratory of Hydroscience and Engineering, Tsinghua University, Beijing
100084, China.

**Jian Sun**

State Key Laboratory of Hydroscience and Engineering, Tsinghua University, Beijing
100084, China.

**Binliang Lin**

State Key Laboratory of Hydroscience and Engineering, Tsinghua University, Beijing
100084, China.

**Abstract**: A function for the bed-load sediment transport rate is derived. This is achieved from the first principle by using the entrainment probabilities of the sediment grains by rolling and lifting, and by introducing two travel lengths, respectively, for the first time. The predictions from the new bed-load function agree well with the experimental results over the entire experimental range and show significant improvement over the commonly used formula for bed-load transport rate. The new function shows that, in terms of contributing to the bed-load transport rate, the total entrainment probability of the sediment grains is a weighted summation of those by the lifted and rolling grains, rather than a simple addition of the two. The function has also been used to predict the total entrainment probability, saltation length and the bed layer thickness at high bed-load transport rate. These predictions all agree well with the experimental results. It is found that, on average, the travel length for the rolling sand grains is about one order of magnitude less than that for the lifted ones.

1. **Introduction**

Sediment transport is part of the fluvial processes and is closely related to the morphological changes. In general, there are two forms of sediment transport, which are called the bed load and suspended load. In bed load, the sediment grains may slide, roll or travel in succession of low jumps, termed saltation, but close to the bed surface from where they may leave temporarily (Dey, 2014).

It is generally believed that the bed load is closely related to the entrainment of the sediment grains being pickup by the fluid motion near the bed surface when the hydraulic force acting on the grain is larger than the grain resistance. Here pickup means that the sand grains have been moved relative to the stationary bed surface. The movement can be in the modes of rolling, sliding, bouncing and lifting. Over the years, the pickup of the sediment grains by the fluctuating motions of the flow near the bed surface has been extensively investigated theoretically, experimentally and numerically (Leighly, 1934; Einstein and El-Sammi, 1949; Sutherland, 1967; Paintal, 1971; Grass, 1970, 1983; Apperley and Raudkivi, 1989; Jain, 1991; Nelson et al, 1995; Cao, 1997; Papanicolaou et al, 2001; Nelson et al 2001; Sumer et al, 2003; Zanke, 2003; Paiement-Paradis et al, 2003; McEwan et al, 2004; Hofland et al, 2005; Schmeeckle et al, 2007; Vollmer and Kleinhans, 2007; Dwivedi et al, 2010, 2011; Valyrakis et al, 2010; Celik et al, 2010; Valyrakis et al, 2013; and Agudo et al, 2014). In these investigations, the dependence of the incipient motions on the magnitude of instantaneous velocity, the roughness and irregularity of the bed surface texture and morphology, average bursting period and area, impulse (instantaneous forces and duration) and flow power has been studied.

Given the complexity of the sediment transport in the bed layers where there exist complex interactions between a large amount of sediment grains and interactions between solid grains and the turbulent flows, it is difficult to predict the bed-load transport rate analytically based on the incipient motions of individual sand grain. So far, there are mainly two approaches in deriving the bed-load formula theoretically. One is the deterministic approach such as those by Bagnold (1954) and Yalin (1977). The second is the probabilistic approach such as those by Einstein (1942, 1950), Engelund and Fredsoe (1976), Wang et al. (2008) and Zhong et al (2012). The estimated bed-load transport rate from many of the bed-load formulae in general compare well with the experimental results in a limited range and it has been a challenge to develop formulae to predict results with good agreement covering all the experimental range.

In this work, we will take the probabilistic approach since all the investigations on the instantaneous velocity, average bursting period and area, impulse (instantaneous forces and duration) and flow power show that the pickup of the sediment grains from the bed surface is related to the instantaneous velocity (mean plus fluctuations) of the flow. Given that the flows in nearly all the engineering interested open channels with sediment transport are turbulent, the instantaneous flows (or pressure fluctuations, bursting duration and area, impulse and energy) near the bed surface will be random processes, and thus a probabilistic approach will be appropriate.

Einstein (1942, 1950) is the pioneer to develop a bed-load transport model based on the probabilistic concept. By considering the probability of the dynamic lift on sediment grains being larger than its submerged weight, he was able to derive theoretically a pickup function. By combining this function with a travel length of the moving grains near the bed surface and an assumed time scale for the grains to be lifted, Einstein (1950) was able to derive a bed-load formula (with many parameters determined by experimental results). The advantage of using the probabilistic approach over that of deterministic one is that a threshold criterion for the initiation of the grain movement is avoid since it is always a difficult proposition to define (Dey, 2014). The pickup probability function of Einstein (1950) was based on the instantaneous velocity instead of the time-averaged velocity. Engelund and Fredsoe (1976) suggested that the pickup probability is related to both the dimensionless bed load and the dimensionless shear stress when considering the motion of the individual sediment grain contributing to the bed load. They determined their probability function by empirically fitting the experimental data. Sun and Donahue (2000) derived a bed-load formula for the arbitrary size fraction of non-uniform sediment grains based on a method of combining stochastic processes with mechanics. They considered both the probability of fractional incipient motion and that from the average velocity of grain motion.

Cheng and Chiew (1998) investigated the effect of lift coefficient on the pickup probability in a hydraulically rough flow and concluded that the pickup probability is about 0.6% when compared to the Shields criterion for sediment transport. In Cheng and Chiew (1998), the probability distribution (pdf) of the instantaneous streamwise velocity is assumed to be Gaussian, the same as that used in Einstein (1950). Instead, Wu and Lin (2002) assumed that the pdf of the instantaneous streamwise velocity is log-normal and Bose and Dey (2013) used a two-sided exponential pdf. They consider only the entrainment probability of the lifted sediment grains.

Papanicolaou et al (2002) developed a quantitative model for predicting the commencement (including both rolling and lifting) of sediment entrainment under three representative bed packing densities corresponding to the isolated, wake interference and skimming flow regimes. Wu and Chou (2003) derived both the rolling and lifting probabilities of the sediment entrainment and assumed that the numbers of sediment grains in sliding and bouncing are small in comparison with those in rolling and lifting. Wu and Chou (2003) considered both the fluctuations of turbulent flow and the randomness of the bed grain geometry. They noted that, in predicting the bed-load transport rate, many of the existing stochastic models such as those from Einstein (1950), Paintal (1971) and Sun and Donahue (2000) are based on a single entrainment mode and have displayed considerable errors, and suggested that a significant improvement may be expected if a more reliable entrainment probability is incorporated into the stochastic modelling.

In this work, following Einstein (1950), we consider the equilibrium condition of exchange of the sand grains between the bed layer and the bed surface. For each unit of time and bed surface area the same number of a given type and size of grains must be deposited in the bed as are scoured from it. Under this condition, the sand grains can be considered as loosely packed on the surface and the flow is hydraulically rough. We will focus on deriving a new function for the bed-load transport rate using the probabilities from both the rolling and lifting sand grains. However, our method of using the pickup probabilities to derive the bed-load function is different from that of Einstein (1950).

We will choose the simple forms of these probabilities to derive the bed-load function. The entrainment probabilities for the rolling and lifting grains will be presented first. This is partially for the completion of the paper. We assume that the pdf of the instantaneous streamwise velocity follows the Gaussian distribution rather than log-normal and the randomness of the bed grain geometry can been taken into account through the averaged acting distances of the lift and drag forces. The derived probability for the lifted grains is similar to that of Cheng and Chiew (1998) but the probability for the rolling grains is given in its current form for the first time.

In order to derive the bed-load transport rate across a unit width (which is a mass flow rate in the flow direction and parallel to the bed surface) from the entrainment flow rate (which is a mass flow rate in the normal to the bed surface direction), transport lengths of the sediment grains before they stop rolling or fall back to the bed under gravity are required. For the lifted sand grains, this distance is called the saltation length. Instead of assuming a single travel length for both the rolling and lifting sediment grains near the bed surface, we use two different travel lengths. This is based on the considerations that the distance travelled by the lifted sand grains is mainly controlled by the flow above the bed surface while the distance travelled by the rolling grains will be affected by both the flow near the bed surface and the roughness of the bed surface (Jain, 1991; Papancalaou et al, 2001). The entrained sediment grains for both the rolling and lifting are then determine separately and the total bed-load transport rate is the sum of these two. The new formula is then compared with the experimental data and those from some commonly used formulae over the entire range of the experimental data. The total entrainment probability, the saltation length and the bed layer thickness at high transport rate are also predicted and compared with the experimental results.

2. **The Probabilities of the Sediment Grains by Lifting and Rolling**

We determine the probabilities from the rolling and lifting grains near the bed surface following the analysis of Wu and Chou (2003) with some simplification. Figure 1 shows the 2D projection of three representative-spherical sediment grains with uniform diameter $d$ and relative positions of the forces acting on the top grain due to the lift $F_L$, drag $F_D$ and the submerged weight $W$.

These forces can be estimated as

$$F_L = C_L \frac{\rho_f A u^2}{2} \tag{1}$$

$$F_D = C_D \frac{\rho_f A u^2}{2} \tag{2}$$

$$W = g(\rho_s - \rho_f)\frac{\pi d^3}{6} \tag{3}$$

where $\rho_f$ and $\rho_s$ are the densities of the fluid and sand grains, respectively, $A$ is the frontal area of the grains, $C_L$ and $C_D$ are the lift and drag coefficients, respectively, $g$ is the gravity acceleration and $u$ is the instantaneous effective velocity, i.e. the representative velocity over the frontal area of the spherical sand grain at a particular instant of time and should fluctuate with time. This is different from the local mean velocity of the flow (which is normally a time average). The three forces $F_L$, $F_D$ and $W$ acting on the spherical grain 1 through the pivotal arms $R_L$, $R_D$ and $R_W$, respectively, about the common contact point $C$ (see Figure 1). Here, we assume that $R_L = R_W$. Instead of determining the three pivotal lengths by considering the randomness of bed geometry as that in Wu and Chou (2003), we will use the averaged $R_L$ and $R_W$ (it is the ratio of $R_D/R_L$ that needs to be considered).

Rolling and lifting of the sediment grain from its resting position occur only when the moments generated by the drag and lift forces about the contact point $C$ (see Figure 1) overcome the moment due to the submerged weight of the grain. According to this, rolling occurs when

$$F_D R_D + F_L R_L > W R_W \quad (4)$$

and lifting occurs when

$$F_L > W \quad (5)$$

Combining (1-3) and (4-5), the thresholds for rolling can be found as

$$u > B_R \quad (6)$$

with

$$B_R = \sqrt{\frac{4}{3} \frac{d \Delta g}{C_D (R_D / R_L) + C_L}} \quad (7)$$

and that for lifting can be found as

$$u > B_L \quad (8)$$

with

$$B_L = \sqrt{\frac{4}{3} \frac{d \Delta g}{C_L}} \quad (9)$$

where $\Delta = (\rho_s - \rho_f)/\rho_f$ is the submerged relative density of the sediment grains. In deriving (7) and (9), we have used the frontal area $A = \pi d^2 / 4$ of a sphere as the area of the drag and lifting forces acting upon the grains. In a complex rough sand surface, some sand grains will be behind others and the effective acting area of the drag and lift forces will be less than this. In Wu and Chou (2003), this variation of frontal area was taken into account by changing the position of the top sphere in Figure 1 relative to others, while we account this by the resulting average arm lengths. Also, we will be using the lift coefficient from the experimental results of Einstein and El-Sammi (1949), which was measured based on the full frontal area of spherical particles.

Comparing the above two thresholds, since $C_D(R_D/R_L) > 0$, it can be concluded that

$$\frac{B_R}{B_L} < 1 \qquad (10)$$

which means that the entrainment threshold for lifting is higher than that for rolling. This is the result given by Wu and Chou (2003) and is consistent with the finding of Ling (1995). Valyrakis et al (2013) arrived at the same conclusion from the energy required to move the sand by rolling and lifting.

The lifting and rolling probabilities of the sediment grains near the bed surface can be determined as

$$P_L = P(u > B_L) \qquad (11)$$

and

$$P_R = P(u > B_R) - P(u > B_L) \qquad (12)$$

In order to determine these probabilities, the pdf of the instantaneous streamwise velocity is required. Cheng and Chiew (1998) used the normal or Gaussian pdf, Wu and Lin (2002) and Wu and Chou (2003) used the log-normal pdf, and Bose and Dey (2013) used the two-sided exponential pdf. Wu and Lin (2002) and Bose and Dey (2013) compared the effect of assuming different pdfs and it is not conclusive whether the non-Gaussian pdfs of the instantaneous streamwise velocity have improved the agreement between the predictions and the experimental results from Guy et al (1966) and Luque (1974). For simplicity, we will use the Gaussian pdf for the instantaneous streamwise velocity centred at the local mean velocity $u_b$, i.e.

$$f(u) = \frac{1}{\sqrt{2\pi}\sigma_u} \exp\left(-\frac{(u-u_b)^2}{2\sigma_u^2}\right) \tag{13}$$

where $\sigma_u = \sqrt{\overline{(u-u_b)^2}}$ is the r.m.s. of the velocity fluctuations.

For turbulent flows above the bed surface, the velocity $u_b$ is located at $y_b$ above the bed surface (see Figure 1). The mean approaching velocity $u_b$ is determined as Cheng and Chiew (1998) and Wu and Lin (2002). It is assumed that the velocity close to the sediment grains next to the bed surface follows the universal logarithmic profile

$$\frac{\bar{u}}{u_*} = \frac{1}{\kappa}\ln\left(\frac{y}{y_0}\right) \tag{14}$$

where $\bar{u}$ is the local mean velocity at distance $y$ from the bed surface, $u_*$ is the wall shear velocity, $\kappa \approx 0.4$ is the von Karman constant and $y_0$ is the zero velocity level (see Figure 1) when assuming a logarithmic velocity profile near the bed surface, and is set to $0.033k_s$ above the bed surface. Here $k_s$ is the height of the representative bed roughness. Cheng and Chiew (1998) considered the most stable situation when the grains rest in an interstice formed by the top layer grains and found that $y_b=0.6d$. Assuming that $k_s = 2d$ and using (14), the mean effective velocity (in terms of producing the drag force $F_D$) approaching a sand grain can be determined to be $u_b = 5.52u_*$.

Combining (11-12) and (13), we obtain

$$\begin{aligned} P_L &= \int_{B_L}^{\infty} f(u)du = 1 - \left[\int_{-\infty}^{u_b} f(u)du + \int_{u_b}^{B_L} f(u)du\right] \\ &= 0.5 - \int_{u_b}^{B_L} f(u)du \end{aligned} \tag{15}$$

and

$$P_R = \int_{B_R}^{B_L} f(u)du = \int_{u_b}^{B_L} f(u)du - \int_{u_b}^{B_R} f(u)du \tag{16}$$

Using the approximation of the error function by Guo (1990), the above probabilities can be approximated as

$$P_L = \frac{1}{2}\{1 - \frac{B_L - u_b}{|B_L - u_b|}\sqrt{1 - \exp[-\frac{2(B_L - u_b)^2}{\pi\sigma_u^2}]}\}  \qquad (17)$$

and

$$P_R = \frac{1}{2}\{\frac{B_L - u_b}{|B_L - u_b|}\sqrt{1 - \exp[-\frac{2(B_L - u_b)^2}{\pi\sigma_u^2}]} - \frac{B_R - u_b}{|B_R - u_b|}\sqrt{1 - \exp[-\frac{2(B_R - u_b)^2}{\pi\sigma_u^2}]}\}  \qquad (18)$$

Based on the experimental results of Kironoto and Graf (1994) from rough boundary open channel flows, Cheng and Chiew (1998) assumed $\sigma_u = 2u_*$. With this approximation, $u_b = 5.52u_*$, and the expressions for $B_R$ and $B_L$ in (7) and (9), respectively, the above probabilities can be expressed as

$$P_L = \frac{1}{2}\{1 - \frac{0.21 - \sqrt{\theta C_L}}{|0.21 - \sqrt{\theta C_L}|}\sqrt{1 - \exp[-(\frac{0.46}{\sqrt{\theta C_L}} - 2.2)^2]}\}  \qquad (19)$$

and

$$P_R = \frac{1}{2}\left[\begin{array}{l} \frac{0.21 - \sqrt{\theta C_L}}{|0.21 - \sqrt{\theta C_L}|}\sqrt{1 - \exp[-(\frac{0.46}{\sqrt{\theta C_L}} - 2.2)^2]} - \\ \frac{0.21 - \sqrt{\theta[C_D(R_D/R_L) + C_L]}}{|0.21 - \sqrt{\theta[C_D(R_D/R_L) + C_L]}|}\sqrt{1 - \exp[-(\frac{0.46}{\sqrt{\theta[C_D(R_D/R_L) + C_L]}} - 2.2)^2]} \end{array}\right]  \qquad (20)$$

where $\theta = u_*/\sqrt{\Delta g d}$ is the dimensionless shear velocity (also called Shields parameter).

Equations (17) and (19) are similar to those given by Cheng and Chiew (1998) but (18) and (20) are different from those given in Wu and Chou (2003), since we have used the Gaussian pdf for the random steamwise velocity.

To determine $P_L$ and $P_R$ in the above equations, the lift and drag coefficients $C_L$ and $C_D$ are required together with the ratio $R_D/R_L$. The experimental data of Patnaik et al (1994) show that the ratio $C_L/C_D$ decreases with $R_p$ for 4,000< $R_p$ <60,000 and that, in general, 0.5 < $C_L/C_D$<1.5. As can be seen from (19) and (20), it is the term $C_D(R_D/R_L)$, i.e., the drag coefficient $C_D$ and the moment arm length ratio $R_D/R_L$ together, that make the contribution to the probability of the rolling grains. Using the arm lengths $R_L$ and $R_D$ as given by Wu and Chou (2003), which vary with the relative position of a sediment grain to its neighbouring grains, it can be estimated that $\bar{R}_D/\bar{R}_L \approx 1.43$. Here $\bar{R}_L$ and $\bar{R}_R$ are the averaged moment arm lengths. We will use this averaged ratio in (20) and $C_D=C_L$ as that in Wu and Chou (2003) (with different $C_L$ values). With these assumptions, the rolling probability (20) can be simplified to

$$P_R = \frac{1}{2}\{\frac{0.21-\sqrt{\theta C_L}}{|0.21-\sqrt{\theta C_L}|}\sqrt{1-\exp[-(\frac{0.46}{\sqrt{\theta C_L}}-2.2)^2]} - \frac{0.135-\sqrt{\theta C_L}}{|0.135-\sqrt{\theta C_L}|}\sqrt{1-\exp[-(\frac{0.295}{\sqrt{\theta C_L}}-2.2)^2]}\} \quad (21)$$

Figure 2 shows the entrainment probabilities for the lifting and rolling grains versus the dimensionless shear velocity $\theta$ predicted using $C_L=0.178$ and $C_L=0.25$, respectively, and the experimental results from Guy et al (1966) and Luque (1974). Also shown in Figure 2 is the total entrainment probability according to Wu and Chou (2003)

$$P_T = P_R + P_L \quad (22)$$

As pointed by Bose and Dey (2013), currently there is no consensus in the literature on the lift coefficient $C_L$ (the same is also true for $C_D$). In predicting the entrainment probability, Cheng and Chiew (1998) used $C_L=0.25$, Wu and Lin (2002) assumed $C_L=0.21$ while Bose and Dey (2013) adapted $C_L=0.15$. The minimum $C_L$ used by Wu and Chou (2003) is 0.36. The $C_L$ values selected for comparison in Figure 2 with the experimental data are based on the following considerations: (1) $C_L=0.178$ is from the experimental results of Einstein and El-Samni (1949); and (2) $C_L=0.25$ is the same as that used in Cheng and Chiew (1998) since we have assumed that the pdf of the random streamwise velocity is Gaussian, the same as that in Cheng and Chiew (1998) and Einstein (1950).

Figure 2 shows that the probabilities shift towards smaller dimensionless shear velocity $\theta$ and the maximum rolling probability decreases as the lift coefficient increases. The maximum rolling probability is about 0.5 for $C_L=0.178$ and about 0.4 for $C_L=0.25$. These are higher than 0.25 as that given in Wu and Chou (2003). Figure 2 also shows that for $C_L=0.178$, the lifting probability is about *0.9* at $\theta=0.8$ and it is higher than *0.97* when $\theta>2$, which means that nearly all the sediment grains at the top of the bed surface can be lifted at such high dimensionless shear velocity (the probability of the rolling sand grains is close to zero). It can be seen from Figure 2 that the experimental results for the entrainment probability at low $\theta$ in general fall between the lifting probabilities of $C_L=0.178$ and $C_L=0.25$, and the predicted total entrainment probability $P_T$ as according to (22) are in general much higher than the experimental data. Wu and Chou (2003) attempted to match the predicted probability of the lifted sediment grains with the experimental data. It is possible that the experimental data on the entrainment probability of the sediment grains would include both the probabilities of that from the lifting grains and that from the rolling grains, and will discuss this further in connection with the experimental results for the bed-load transport rate.

**3. Bed Load Transport Function Based on the Lifting and Rolling Probabilities**

We consider statistically steady flows and assume that the bed load has reached equilibrium. Let's consider an area $A_1$ on the bed surface from where $n$ sediment grains are entrained into the bed layer during a time period $\delta t$. The mass entrainment rate by rolling and lifting can then be calculated from the first principle as

$$\dot{m} = \dot{m}_R + \dot{m}_L = \frac{n}{\delta t} \rho_s \frac{\pi}{6} d^3 \tag{23}$$

where $\dot{m}_R$ and $\dot{m}_L$ are the mass flow rates entrained from the rolling and lifted grains, respectively. As mentioned earlier, we neglect the sediment grains entrained by sliding and bouncing and assume that the number of sediment grains, $n$, entrained into the bed-load layer include those from rolling and lifting only so

$$n = n_R + n_L \tag{24}$$

and $n_R$ and $n_L$ are related to the corresponding rolling and lifting probabilities $P_R$ and $P_L$ by

$$n_R = \frac{A_1}{\pi d^2/4} P_R C_b \tag{25}$$

and

$$n_L = \frac{A_1}{\pi d^2/4} P_L C_b \tag{26}$$

where $P_L$ and $P_R$ are given by (19) and (21) and $C_b$ is the sediment concentration at the bed surface to take into account the availability of the sediment grains at the bed surface (Church, 1978; Cao, 1997). The factor $A_1/(\pi d^2/4)$ in (25-26) is proportional to the number of sediment grains over the surface area $A_1$ assuming that the grains are closely packed, and $C_b A_1/(\pi d^2/4)$ is the average number of sediment grains over the surface $A_1$ available to be picked up. Thus the mass flow rates over the area $A_1$ from the rolling and lifted grains are, respectively,

$$\dot{m}_R = \frac{n_R}{\delta t} \rho_s \frac{\pi}{6} d^3 = \frac{\rho_s}{\delta t} \frac{\pi}{6} d^3 \frac{A_1}{\pi d^2/4} P_R C_b = \frac{2A_1}{3\delta t} d\rho_s P_R C_b \tag{27}$$

and

$$\dot{m}_L = \frac{n_L}{\delta t} \rho_s \frac{\pi}{6} d^3 = \frac{\rho_s}{\delta t} \frac{\pi}{6} d^3 \frac{A_1}{\pi d^2/4} P_L C_b = \frac{2A_1}{3\delta t} d\rho_s P_L C_b \qquad (28)$$

This method of calculating the mass flow rates based on the probability is similar to that used in Engelund and Fredsøe (1976).

To convert the above entrainment rates at the bed surface into the bed-load transport rate $q_b$ across the channel (defined as the volume flow rate of the sediment grains per unit channel width), we need the surface area $A_1$ which can be calculated as

$$A_1 = 1 \times L \qquad (29)$$

where $L$ is the travel length of the grains in the mean flow direction. This is the saltation length for the lifted grains and the rolling length for the rolling grains. Einstein (1950) considered the lifted grains only and assumed that this length is a fixed ratio of the diameter of the sediment grains, i.e. $L=100d$. Based on the experimental data and theoretical considerations, Yalin (1972) and Wang *et al* (2008) suggested that this travel length should depend on the ratio between the average hydrodynamic force on a grain and its submerged weight, i.e.

$$\frac{L}{d} = \lambda \frac{\rho_f u_b^2 d^2}{g(\rho_s - \rho_f)d^3} = \lambda (\frac{u_b}{u_*})^2 \frac{\rho_f u_*^2 d^2}{g(\rho_s - \rho_f)d^3} = \lambda (\frac{u_b}{u_*})^2 \theta \qquad (30)$$

where $\lambda$ is a constant. However, the travel lengths by the rolling and lifted grains should be different so we assume separately that

$$\frac{L_L}{d} = \lambda_1 \theta \qquad (31)$$

and

$$\frac{L_R}{d} = \lambda_2 \theta \qquad (32)$$

As discussed by van Rijn (1984), $L_L$ depends on the height of the grains being lifted. On the other hand, $L_R$ depends on the roughness of the bed surface since the rolling grains, by definition, are in contact with the bed all the time (Papanicolaou et al, 2002). Since the characteristics of the motions by the lifted grains and the rolling grains are different, we believe that using two separate travel lengths for the two modes of sediment transport is appropriate.

For the time scale $\delta t$, again we follow that of Yalin (1972) and Wang et al (2008) and assume that it is related to the grain diameter $d$ and the wall shear velocity $u_*$, i.e.

$$\delta t = \beta \frac{d}{u_*} \qquad (33)$$

Here $\beta$ is a time constant which may depend on the Reynolds number as suggested by Yalin (1972). The above equations (27-28) are similar to that of Cao (1997) except that Cao (1997) used the averaged turbulent bursting period for the time scale and the mean area of the bursting event per unit bed-surface area. By taking into account the average bursting period and area, the pickup probability of the grains from the bed surface has been implicitly included since the mean bursting period and mean area occupied by turbulent bursting used in Cao (1997) are statistical averages.

By combining (27-28) and (31-33), it can be derived that the bed-load transport rates for the lifted and rolling grains are, respectively,

$$\frac{q_{bL}}{\sqrt{\Delta g d^3}} = \frac{2\lambda_1}{3\beta} \frac{u_* \theta}{\sqrt{\Delta g d}} P_L C_b = \frac{2}{3\beta} \lambda_1 \theta^{3/2} P_L C_b \qquad (34)$$

and

$$\frac{q_{bR}}{\sqrt{\Delta g d^3}} = \frac{2\lambda_2}{3\beta} \frac{u_* \theta}{\sqrt{\Delta g d}} P_R C_b = \frac{2\lambda_2}{3\beta} \theta^{3/2} P_R C_b \qquad (35)$$

Adding these together gives the total bed-load transport rate from both the lifted and rolling grains as

$$\Phi = \frac{q_b}{\sqrt{\Delta g d^3}} = \frac{q_{bL} + q_{bR}}{\sqrt{\Delta g d^3}} = \frac{2\lambda_1}{3\beta}(P_L + \alpha P_R)\theta^{3/2} C_b$$
$$= B(P_L + \alpha P_R)\theta^{3/2} \tag{36}$$

Here $B = 2\lambda_1 C_b/(3\beta)$ is a constant and $\alpha = \lambda_2/\lambda_1$ is the ratio of the travel length by the rolling and lifted sand grains.

In (30-36), the constants $\beta, \lambda_1, \lambda_2$ are introduced to help the flow of the derivation. The physical meaning of $\beta$ is the dimensionless time constant and the values of $\lambda_1$ and $\lambda_2$ will be given in Section 4.2. Eq. (36) shows that, as in many of the existing formulae, the new bed-load function has only two constants, $B$ and $\alpha$. These need to be determined from the experimental results.

Figure 3 shows the calculated results using (36) with $B=20$ and $\alpha = 0.15$, i.e.

$$\Phi = 20(P_L + 0.15 P_R)\theta^{3/2} \tag{37}$$

together with the experimental results from Einstein (1942), Peter-Meyer and Muller (1948), Smart (1984) and Wilcock (1988). The lift coefficient $C_L = 0.178$ from Einstein and El-Samni (1949) has been used. Also shown in Figure 3 are the calculated results from (36) with $\alpha = 0$ and $\alpha = 1$ (both with $B=20$), and the bed-load transport rates calculated using formulae from Peter-Meyer and Muller (1948), Einstein (1950), Nielsen (1992) and Cheng (2002). In applying (37), (19) and (21) are used in calculating the probabilities $P_L$ and $P_R$.

The results in Figure 3 shows that, at low $\theta$, the present results using the lifting probability $P_L$ will in general under-estimate the bed-load transport rate and that based on $P_T$ as given by (22) will in general over-estimate it. Figure 3 shows that the results calculated using (37) agree with the experimental results the best over the entire range, i.e. $10^{-5} < \Phi < 100$. In contrast, the results from Peter-Meyer and Muller (1948), Einstein (1950), Nielsen (1992) and Cheng (2002) are in general under-estimating the bed-load transport rate in either the low $\theta$, high $\theta$, or both, range (s).

Eqn. (37) shows that, at high $\theta$, the dimensionless bed-load transport rate follows the well-known $\Phi \propto \theta^{1.5}$ law as that given by many of the existing formulae. However, for $\Phi < 1$, the effect from the probability functions $P_L$ and $P_R$ on the bed load increases as $\theta$ decreases. Figure 3 also shows that for $\Phi > 10^{-2}$ the bed load transport rate is dominated by lifted sand grains while for $\Phi < 10^{-2}$, the bed load is dominated by the rolling sand grains. This is consistent with the suggestions by van Rijn (1987) and Dey (2014).

A close look at the derivation resulting in (37) reveals that the factor $\Phi \propto \theta^{1.5}$ comes from the assumed travel lengths of the sediment grains as given by (31-32) and the time scale as given by (33) since the term $P_L + \alpha P_R \approx 1$ at high $\theta$. As pointed out by Laursen (1999), the problem with Einstein's derivation is that "even at large values of lift, the probability could not become greater than unity, and the rate of the movement stopped increasing when the turbulence effect was lost". The reasons for this problem are that the length scale introduced by Einstein (1950) for the saltation length is fixed at 100 times of the sand diameter and does not depend on the flow (such as the dimensionless shear velocity) and the time scale he introduced depends only weakly on the flow. The travel lengths introduced in (31-32) depend on the dimensionless shear velocity which means that at high $\theta$, the sand grains pickup by the flow (resulting in either the rolling motion or saltation, but mainly the sand grains lifted) will travel further distance before they fall back to the bed surface than that at low $\theta$. Thus the $\Phi \propto \theta^{1.5}$ in (36) depends on the dimensionless shear velocity $\theta$ which is a mean flow quantity not affected by turbulent fluctuations and is thus independent of the forms of the pdf used for velocity fluctuations.

Einstein (1950) derived the bed-load function using the probability of the lifted sand grains only. The present bed-load function is derived from probabilities of rolling and lifted sand grains. Thus the starting point of the present derivation is similar to that of Einstein (1950). However, the present method of using the probabilities to derive the bed-load function is different from that of Einstein (1950). While Einstein's results under-predict the bed-load transport rate at high dimensionless shear velocity (see Figure 3), (37) is able to predict the transport rate over the entire experimental range. The results in Figure 3 seems to support partially the suggestion of Cheng and Chiew (1999) that there are limitations in Einstein's derivation of the bed-load function but the original definition of the pickup probability should not pose any major problems in relating the initial motion of the sand grains to the near-bed turbulent velocity.

## 4. Applications

**4.1 Total entrainment probability**

A close look at (37) shows that, in determining the bed-load transport rate, a proper account of the total entrainment probability from both the rolling and lifting sediment grains is not by simply adding the entrainment probabilities from the lifting and rolling sediment grains (see (22)), rather, it should be calculated as

$$P_E = P_L + 0.15 P_R \tag{38}$$

Wu and Chou (2003) assumed that the rolling and lifting are mutually independent and thus suggested that the mean total probability of entrainment should be calculated using (22). The derivation of (36) suggested that, in determining the bed-load transport rate, using (22) is equivalent to assuming that the average travel length of the rolling sand grains is the same as that of the lifted grains, i.e. $L_R = L_L$. Results in Figure 3 clearly show that using the total probability of entrainment as given in (22) will grossly over-estimate the sediment transport rate at low dimensionless shear velocity $\theta$ (the case with $\alpha =1$) and thus disrepute such an assumption.

Figure 4 shows the total entrainment probability $P_E$ calculated using (38) and the experimental results of Guy et al (1966) and Luque (1974). Comparing the results from Figure 4 and those from Figure 2, $P_E$ given in Figure 4 agrees with the experimental results much better than any of the predictions based on either $P_L$ or $P_T$ alone as shown in Figure 2. This suggests that the experimental results on the entrainment probability as presented in Figure 4 may have included both the lifted and rolling sediment grains. However, a proper account on their contribution to the sediment transport rate is that given by (38) rather than that by (22). It should be pointed out that, in comparing the prediction of the entrainment probability using (38) with the experimental results in Figure 4, we have not adjusted any of the given constants $B$ and $\alpha$ further. These are determined from the comparison with the bed-load transport rate as shown in Figure 3. It should also be mentioned that the total entrainment probability as given by (38) is a prediction from (37).

**4.2 Grain travel length**

Since $\alpha = \lambda_2 / \lambda_1 = L_R / L_L$ is the ratio of the travel lengths by the rolling and lifted grains, the results from Figures 3 and 4 with $\alpha = 0.15$ indicates that, in bed load, the travel length of the rolling rains is close to an order of magnitude smaller than that of the lifted grains. This will have consequence in determining the adaption length in simulating non-equilibrium sediment transport (Philip and Sutherland, 2000). In simulating the sediment transport with non-equilibrium bed load, an adaption length $L_x$ is generally required as the flow or bed conditions changes. It has been generally assumed that $L_x$ is closely related to $L_L$. The present results show that such a single adaption length based on the travel length of the lifted grains may not be appropriate for the rolling grains.

Using $B = 2\lambda_1 C_b/(3\beta) = 20$, $u_b/u_* = 5.52$ as given before, and assuming $C_b = 0.65$ for the sediment grains of uniform diameter at the bed surface (packing density over a surface by spherical sand grains of uniform diameter) in collecting the experimental data shown in Figure 3, it can then be estimated, using (31-32), that

$$\frac{L_L}{d} = \lambda_1 \theta \approx 46.2\beta\theta \tag{39}$$

and

$$\frac{L_R}{d} = \lambda_2 \theta \approx 6.9\beta\theta \tag{40}$$

Equations (39-40) show that $\lambda_1 \approx 46.2\beta$ and $\lambda_2 \approx 6.9\beta$.

The travel length for the lifted grains given in (39) is close to that determined experimentally by Nino et al (1994), which gives $\lambda_b/d \approx 2.3\theta/\theta_c$, where $\lambda_b$ is the saltation length (assuming the critical dimensionless shear velocity $\theta_c = 0.05$ and $\beta \approx 1$). Equations (39-40) show that it is necessary in simulating the non-equilibrium bed-load transport, to separate the contributions from the lifted grains and that from the rolling grains, and apply the respective adaption lengths accordingly. Equations (39-40) only show that there is a large difference between the travel lengths for the lifted and rolling grains and the actual adaption lengths in simulating non-equilibrium bed-load transport may be many times larger than those given by (39-40) as suggested by Philip and Sutherland (2000).

**4.3 Bed-load layer thickness at high bed-load sediment transport rate**

It has been observed (see Wilson, 2005) that, when $\theta > 0.8$, ripples and dunes are smoothed out and the stationary bed is topped by a sheet-flow layer of bed-load grains in intense motion. This layer is associated with very large sediment transport rate and is of great practical importance for natural flows including rivers in floods. Here we will make no distinction between the sheet-flow layer and bed-load layer and consider the sheet-flow layer as the bed-load layer with intense grain motions. Results in Figure 3 show that (37) can still be used to estimate the bed-load transport rate in intense bed flows at $\theta > 0.8$.

It has been demonstrated by many experimental results, such as those from Sumer et al (1996), Pugh and Wilson (1999) and Capart and Fraccarollo (2011) that, in the sheet-flow layer, stationary bed gives away to moving solids and the concentration of the sediment near the bed surface decreases with height linearly, initially, and then non-linearly. The extrapolation of the linear segment intercepts with the vertical ordinate (distance from the stationary bed) produces a clear identification of the top of the sheet-flow layer (See Pugh and Wilson, 1999) and thus gives a clear definition of the sheet-flow layer thickness $\delta_s$. The average concentration of the sediment in the bed layer (in this case, it is the sheet-flow layer) can be estimated, as that given by Einstein (1950), as

$$\overline{C} = \frac{q_b}{u_b \delta_s} \approx \frac{B_1}{(u_b/u_*)} \frac{d}{\delta_s} C_b \theta \tag{41}$$

where $B_1 = B/C_b \approx 30.8$. In (41), we have used the fact that

$$P_L + \alpha P_R \approx 1, \quad \theta > 0.8 \tag{42}$$

As observed by Sumer et al (1996), Pugh and Wilson (1999), Wilson (2005) and Capart and Fraccarollo (2011), the concentration profile of the sediment in the sheet-flow layer is in essence a linear decrease of the solid concentration with height, thus the average concentration in the sheet-flow layer is $\overline{C} \approx 0.5 C_b$ (Wilson, 2005). Combining this with (41) and using $B_1=30.8$ and $u_b/u_* = 5.52$ as given early, it can thus be estimated from (37) that

$$\frac{\delta_s}{d} \approx 11.1\theta \tag{43}$$

This is in close agreement with the sheet-flow layer thickness suggested by Wilson (1987)

$$\frac{\delta_s}{d} \approx 10\theta \quad (44)$$

Sumer et al. (1999) have also measured the sheet-flow thickness. Using linear regression (with zero interception), the experimental results of Sumer et al. (1999) can be approximated as

$$\frac{\delta_s}{d} \approx 12.4\theta \quad (45)$$

Thus the present prediction (43) falls between the experimental results of Wilson (1987) and Sumer et al (1999).

The bed-layer thickness determined from (43) and the findings of Sumer et al (1999) have direct consequences on the specification of the reference locations in determining the suspended load and thus in calculating the total load at high $\theta$. A proper specification of the reference location and the method of calculating the total load are given in the Appendix.

The constant *B* in (37) is higher than those in the formulae given by Peter-Meyer and Muller (1948), Nielsen (1992) and Cheng (2002). The predicted numerical coefficients for the travel length in (39) and the bed-layer thickness in (43) are directly related to the constant *B*. The fact that the predicted saltation length and bed-layer thickness at high bed-load transport rate agree well with the experimental results seems to support such a high value of *B*.

## 5. Conclusions

- Two travel lengths for the rolling and lifting sand grains in bed layer, respectively, are introduced. The travel length for the rolling grains is introduced for the first time.

- A new bed-load sediment transport rate function has been derived based on the probabilities of the rolling and lifted sediment grains, and the assumed travel lengths and time scale which depend strongly on the flow

- This new function has been compared with the experimental results and agrees well with the experimental results over the entire experimental range.

- In terms of contributing to the bed load, the total entrainment probability of the sand grains is a weighted summation of those from the lifted grains and rolling grains, respectively, rather than a simple addition of the two, or by the lifted sand grains only. The weighting is related to the ratio of the two travel lengths.

- On average, the travel length of the rolling sand grains is about one order of magnitude less than that of the lifted ones.

- The saltation length of the lifted sand grains has been predicted using the new bed-load transport rate function and agrees with some of the experimental results.

- The new bed-load function has also been used to predict the bed-layer thickness (sheet-flow layer thickness) at high dimensionless shear stress and the prediction agrees well with the experimental results.

*Acknowledgements*

**Appendix: Total sediment transport rate**

Sumer et al. (1999) found that, by using the experimental results at $\theta = 4.5$, the measured sediment concentration profile above the bed-load layer agrees well with the familiar Rouse (1937) distribution, namely,

$$C_v = C_a \left( \frac{H-y}{y} \frac{y_a}{H-y_a} \right)^{\frac{w_s}{\kappa u_*}} \tag{A1}$$

where $H$ is the channel depth, $y_a$ is the reference location, $C_a$ is the volumetric concentration of the sediment at $y_a$, and $w_s$ is the terminal velocity of the sediment grains. Their comparison was made at the Rouse parameter $w_s/\kappa u_* = 0.3$. They also found that, as long as the reference location satisfies $0.5\delta_s \leq y_a < \delta_s$, the measured concentration profile agrees well with the Rouse distribution given by (A1) when the corresponding concentration $C_a$ at $y_a$ is used.

Based on these findings, it is recommended that, when the estimation of suspended concentration at high dimensionless shear velocity is required using (A1), $C_a$ can be calculated using (41), with $\delta_s$ from (43), the reference location $y_a$ can be set at

$$y_a = 0.5\delta_s \tag{A2}$$

This is different from that used by Einstein (1950) ($y_a = 2d$) or $y_a = 0.05H$ used in some engineering applications.

The advantage of using (A2) is that the reference location depends on $\theta$. The suspended load $q_s$ (defined as the volume flow rate of the solid per unit channel width for the sediment) can be calculated as that in Einstein (1950), namely

$$q_s = \int_{y_a}^{H} C_v \bar{u} \, dy \tag{A3}$$

with $C_v$ from (A1) and $\bar{u}$ from (14). Since the lower limit of integration in (A3) is $y_a$ rather than $\delta_s$, the total sediment transport rate can be calculated as

$$q_t = q_s + 0.56 q_b \tag{A4}$$

to avoid double counting in the layer $0.5\delta_s < y < \delta_s$ ($q_b$ is calculated from $0 < y < \delta_s$). The factor 0.56 before $q_b$ in the above equation is determined from the assumed linear concentration profile near the bed surface in the sheet layer and the velocity profile in the sheet flow layer of that given by Sumer et al. (1999)

$$\frac{\bar{u}}{u_*} = 14(\frac{y}{\delta_s})^{3/4} \qquad (A5)$$

with $y_a$ located inside the bed layer as that given in (A2).

**Abbreviations**

$A$ = frontal area of a sediment grain (m²)

$A_1$ = Surface area of the bed surface

$B$ = constant

$B_L$ = threshold for a sediment grain to be lifted (m/s)

$B_R$ = threshold for a sediment grain to be rolled (m/s)

$\overline{C}$ = average volumetric concentration of sediment grain in the bed layer

$C_a$ = sediment volumetric concentration at the reference location

$C_b$ = sediment volumetric concentration at the bed surface

$C_D$ = drag coefficient

$C_L$ = lift coefficient

$C_v$ = sediment volumetric concentration in the suspended region

$d$ = diameter of the sediment grain (m)

$F_D$ = hydraulic drag force acting on the grain (N)

$F_L$ = hydraulic lift force acting on the grain (N)

$f$ = probability distribution function of the streamwise velocity fluctuation

$g$ = gravity acceleration (m/s²)

$H$ = channel depth (m)

$k_s$ = equivalent roughness of the sediment grain (m)

$L$ = travel length of a sediment grain (m)

$L_L$ = travel length of a sediment grain by lifting (m)

$L_R$ = travel length of a sediment grain by rolling (m)

$L_x$ = adaption length of sediment grain (m)

$\dot{m}$ = mass flow rate of sediment grain from the bed surface (kg/s)

$\dot{m}_L$ = mass flow rate of sediment grain from the bed surface by lifting (kg/s)

$\dot{m}_R$ = mass flow rate of sediment grain from the bed surface by rolling (kg/s)

$n$ = number of sediment grains in motion

$n_L$ = number of sediment grain in motion by lifting

$n_R$ = number of sediment grain in motion by rolling

$P_E$ = total entrainment probability of the sediment grain as according to (40)

$P_L$ = probability of the lifted sediment grains

$P_R$ = probability of the rolling sediment grains

$P_T$ = total probability of the sediment grain in motion as according to (24)

$q_b$ = total transport rate of the bed load (m$^2$/s)

$q_{bL}$ = transport rate of the bed load by lifting (m$^2$/s)

$q_{bLR}$ = transport rate of the bed load by rolling (m$^2$/s)

$q_s$ = transport rate of the suspended load (m$^2$/s)

$q_t$ = total transport rate of the sediment grain (m$^2$/s)

$R_L$ = moment arm length by the lifting force (m)

$R_R$ = moment arm length by the drag force (m)

$R_W$ = moment arm length by the gravity force (m)

$u$ = instantaneous streamwise velocity (m/s)

$\bar{u}$ = average streamwise velocity (m/s)

$u_b$ = average streamwise velocity acting on a sediment grain (m/s)

$u_*$ = wall shear velocity (m/s)

$W$ = submerged weight of a sediment grain (N)

$w_s$ = terminal velocity of the sediment grain (m/s)

$y$ = distance from the bed surface (m)

$y_0$ = distance where local mean velocity is zero (m)

$y_a$ = the reference location in using the Rouse profile for the suspended load (m)

$y_b$ = location where the average velocity is $u_b$ (m)

$\alpha$ = the ratio of the travel lengths of the sediment grains by rolling and that by lifting

$\beta$ = constant

$\Delta$ = submerged relative density of the sediment grains

$\delta t$ = time scale of the sediment grain in motion (s)

$\delta_s$ = thickness of the bed layer (or sheet flow layer) (m)

$\Phi$ = non-dimensional transport rate of the bed load

$\kappa$ = Karman constant

$\lambda$ = constant

$\lambda_1$ = constant

$\lambda_2$ = constant

$\lambda_b$ = saltation length (m)

$\theta$ = shields parameter

$\theta_c$ = critical shields parameter

$\rho_f$ = density of the fluid (kg/m$^3$)

$\rho_s$ = density of the sediment grain (kg/m$^3$)

$\sigma_u$ = standard derivation of the streamwise velocity fluctuation

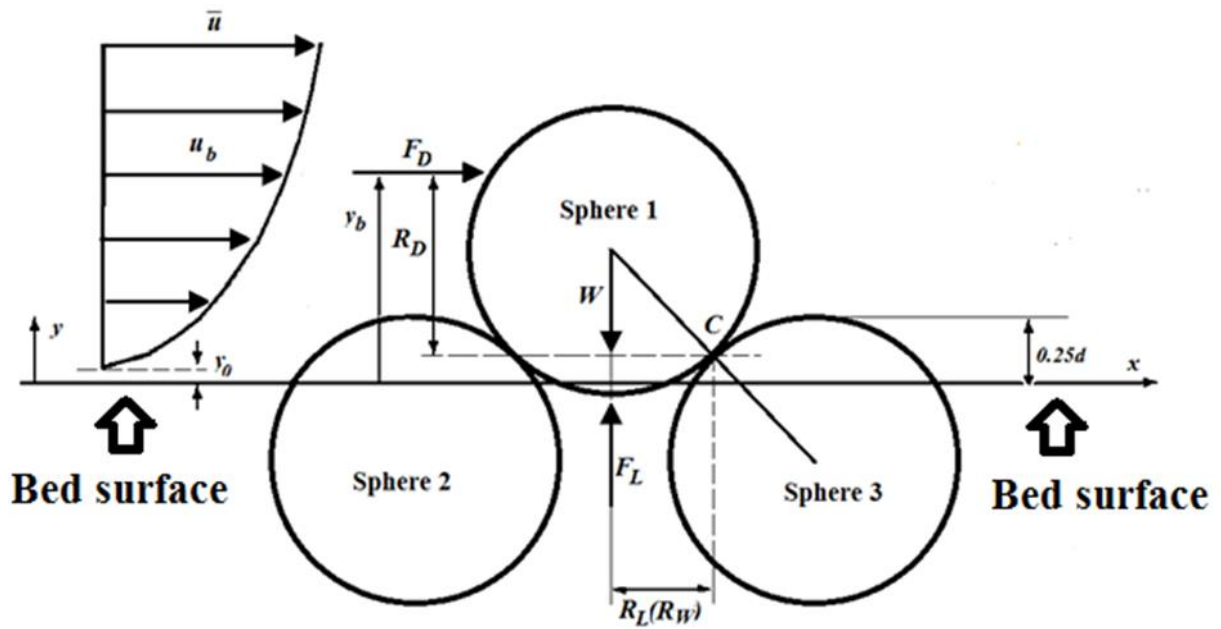

Figure 1 Sketch showing the forces and moment arms acting on sphere 1 together with a mean velocity profile near the bed and several relevant positons. (After Wu and Chou, 2003)

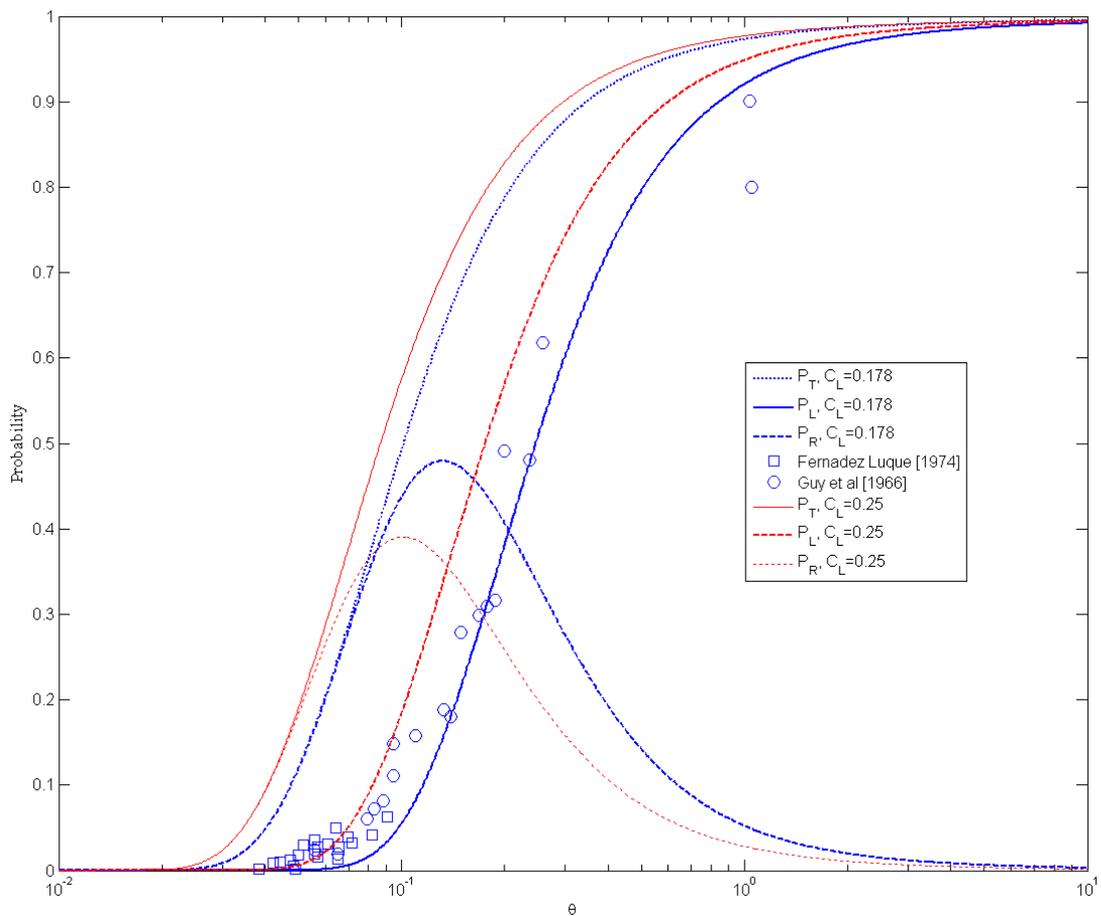

Figure 2 Probabilities from the rolling and lifting grains versus the dimensionless shear velocity for $C_L=0.178$ and $C_L=0.25$. Also shown are the total probabilities as according to (22)

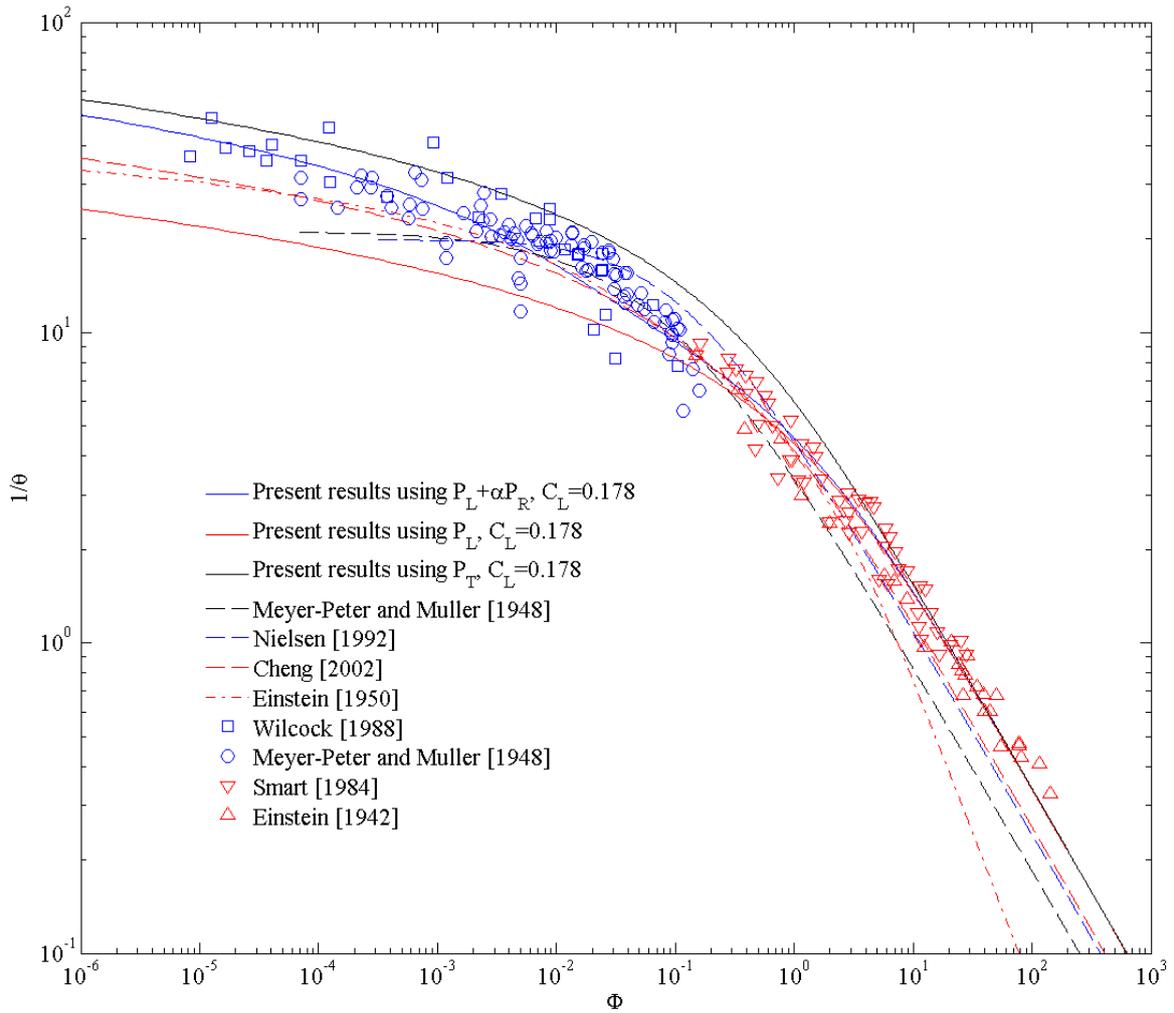

Figure 3 Comparison of the dimensionless bed-load transport rate calculated from (37) with the experimental results and several commonly used formulae. The experimental data are after Wang et al (2008).

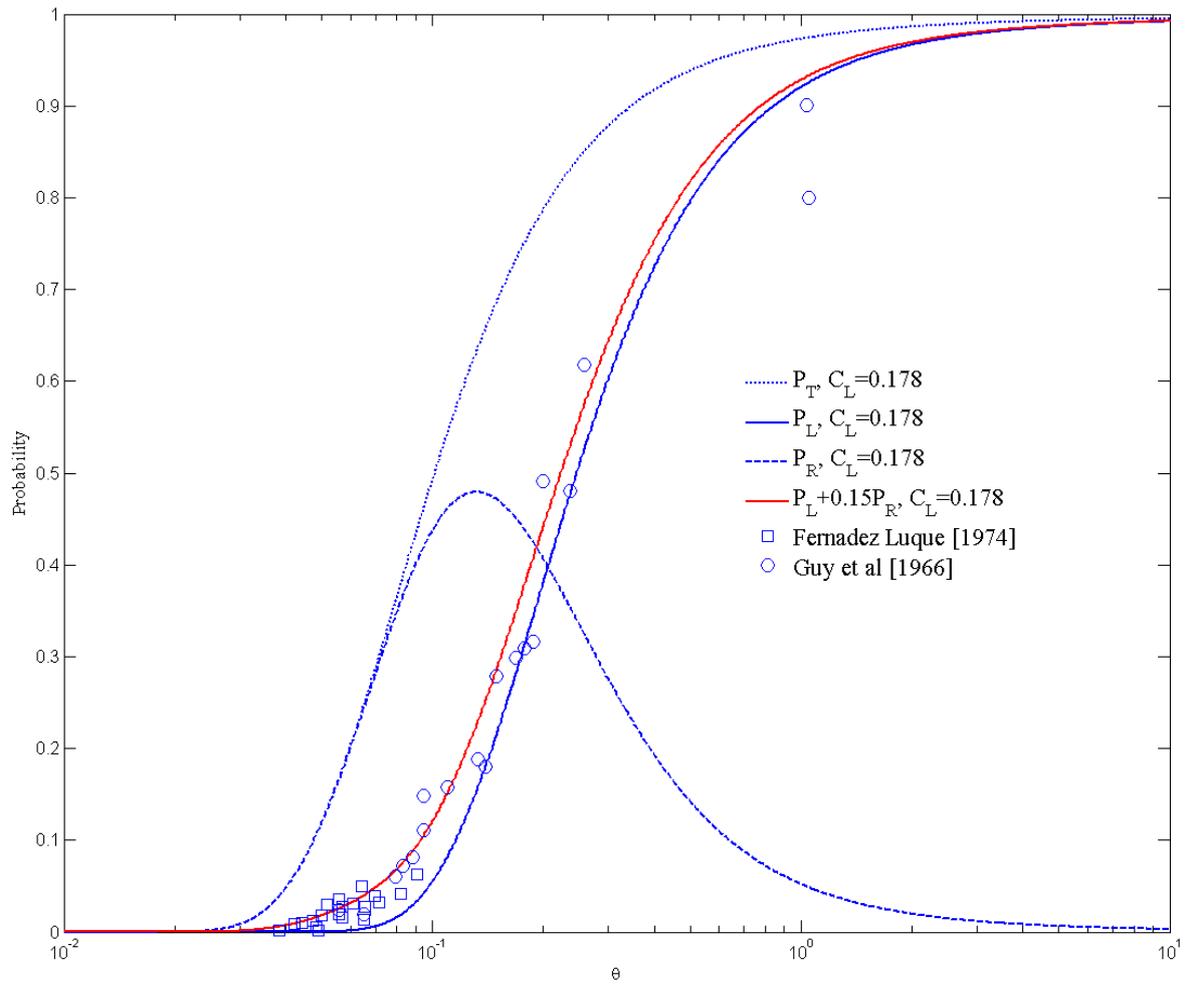

Figure 4 Comparison of entrainment probabilities calculated from (38) with experimental data for $C_L=0.178$.